# Statistics of Local Seismic Emission from the Solar Granulation

**Charles Lindsey**

NorthWest Research Associates, 3380 Mitchell Lane, Boulder, Colorado 80301, USA

E-mail: `clindsey@cora.nwra.com`

**Alina-Catalina Donea**

Center for Astrophysics, School of Mathematical Science, Monash University, Victoria 3800, Australia

**Abstract.** We apply computational seismic holography to high-frequency helioseismic observations of the quiet Sun from *SDO*/HMI to locate predominant sources of seismic emission with respect to the structure of the solar granulation. The regions of greatest seismic emission are the edges of photospheric granules. Seismic emission from regions whose continuum brightnesses are 95-100% of the mean, as resolved by HMI, are about 2.5 times as emissive as regions whose brightnesses are 100-104% of the mean. The greater seismic emissivity from regions whose brightnesses are somewhat less than the mean is roughly in line with expectations from an understanding that attributes most seismic emission to cool plumes plummeting from the edges of granules. However, seismic emission from regions whose continuum brightnesses significantly exceed 104% of the mean is also remarkably high. This unexpected feature of high-frequency seismic emission from the solar granulation begs to be understood.

## 1. Introduction

The discovery of the 300-second oscillations by Leighton, Noyes & Simon [8] and by Evans & Michard [7], and the recognition [11] that these are a manifestation of acoustic waves traveling through the solar interior opened a new era in solar research in the $20^{th}$ century. Since then, our understanding of how these waves are generated has undergone a long evolution. Today, it is generally accepted that the solar oscillations, now popularly expressed as "p-modes", are driven mostly by convection in the Sun's outer convective envelope, hence, closely associated with the solar granulation [15]. Observational evidence of this includes the recognition of conspicuous episodes of outgoing ripples emanating from apparent source regions that tend to be located in intergranular lanes [16].

Since the late 1990s we have developed computational seismic holography for diagnostics of phenomena similar to seismic emission from the quiet Sun but more anomalous, most particularly to transient seismic emission from flares [3, 4] and to acoustic-emission halos surrounding active regions [5]. Recent applications of this by Alvarado Goméz et al. [1] to helioseismic observations by *SDO*/HMI of transient seismic emission from flares suggest that seismic holography in the high-frequency p-mode spectrum can resolve the source distributions of seismic emission sufficiently to locate it statistically with respect to the structure of the solar granulation.[1] This is the object of this study.

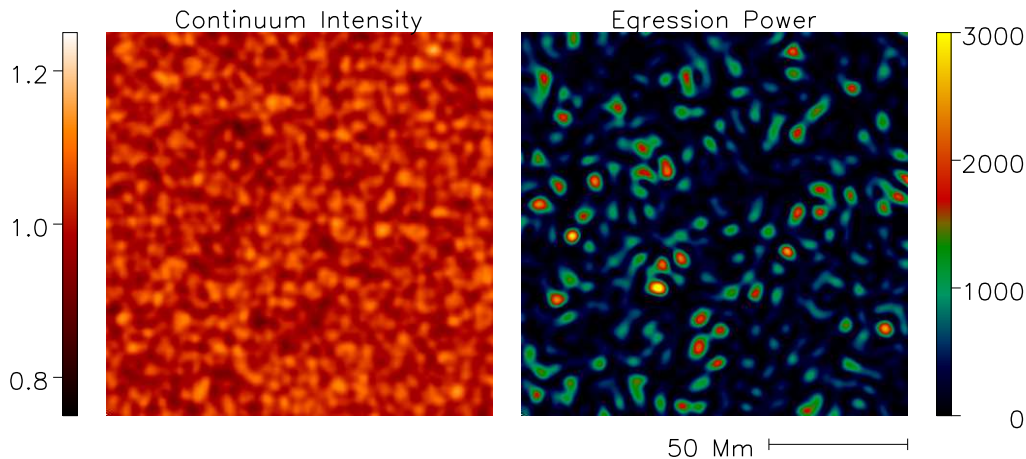

**Figure 1.** Cospatial, cotemporal samples of continuum intensity in the neighborhood of FeI $\lambda 6701$ Å, as resolved by *SDO*/HMI (left), and egression power in the 5.5–6.5-mHz seismic spectrum (right).

## 2. Acoustic-Power Holography of the Solar Granulation
*2.1. Cospatial, Cotemporal Maps of Seismic Source-Power and Continuum Brigtness*

This study shows preliminary results of a comparison between helioseismic source-power maps determined by seismic holography in the 6-mHz spectrum and the structure of the granulation as discriminated in terms of brightness. We apply "subjacent-vantage helioseismic holography" [10] to Doppler observations of the quiet Sun in the 5.5–6.5 mHz spectrum over a square region, $\mathcal{R}$, centered upon the Sun whose area, $A$, is 0.11 $R_\odot^2$ over a period, $T$, of 24 hr. For details on the general technique of seismic holography, we refer to [10]. For details on the application of the technique to episodic phenomena, we refer to [3].

Figure 1 shows cospatial maps of continuum intensity (left) and "egression power" (right), the holographic representation of seismic source-power density, over a small sub-region of $\mathcal{R}$. The egression-power (right) has a spatial resolution of 2.5 Mm, a diffraction limit attained by applying a computational "pupil" (see Fig 4, p. 269 of [1]) whose radial range was 7–28 Mm. This is greater than the $\sim$1.4-Mm cell size of a single granule [2]. However, as the analysis that follows will confirm, the availability of $\sim 5\times 10^6$ granules appearing in $\mathcal{R}$ over $T$ gives us the statistical leaverage us to discriminate mean relative locations of helioseismic signatures to within a small fraction of the diffraction limit.

Subjective statistics obtained by flickering between cospatial pairs of comparative maps such as those shown in Figure 1 show that kernels representing the largest egression powers tend to be centered near the edges of granules, i.e., in regions of intermediate brightness between the centers of granules and the centers of intergranular lanes. These subjective impressions are reinforced by cross-correlation statistics we will now describe.

---

[1] In fact, the solar photosphere is an efficient reflector of low-frequency p-modes [6, 9]. So, apparent sources of low-frequency ($\sim$3 mHz) emission from the neighborhood of any focus are mostly local reflections of p-mode energy generated elsewhere. At high frequencies (e.g., 6 mHz), the photosphere appears to be an efficient absorber of p-modes [6, 9]. So, seismic source signatures at high frequencies are a much stronger representation of acoustic power generated at or near the focus of the diagnostic.

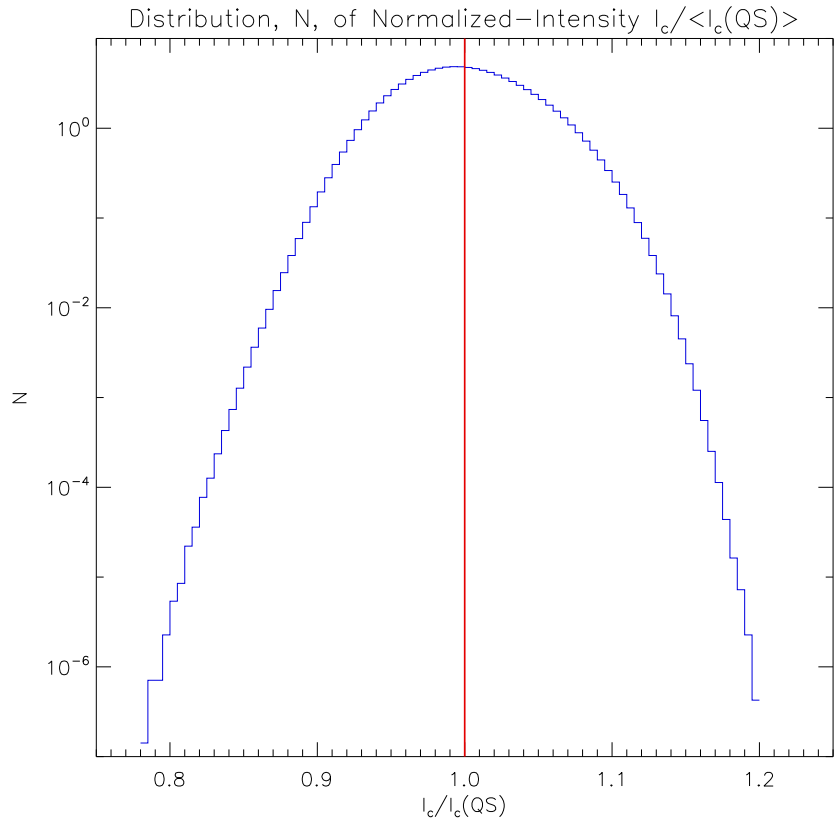

**Figure 2.** Histogram of the distribution of continuum intensity of the quiet Sun, as resolved by SDO/HMI, in the neighbohood of FeI $\lambda$6701 Å, normalized to unity for the mean quiet Sun.

*2.2. Brightness and Seismic Source-Power Statistics*
Figure 2 shows a histogram of the continuum brightness of the quiet Sun normalized to unity over $\mathcal{R}$ for the mean quiet Sun, over the period $T = 24$ hr. This histogram is partitioned into intervals of 0.005 units in quiet-Sun normalized intensity. Figure 3 plots the egression power averaged over the same intervals, over the range 0.85–1.15. This profile shows a strong statistical dependence of the mean egression power upon the local continuum brightness of the solar granulation. To discriminate regions of different seismic emissivity, we recognize three "continuum-brightness regions", I, II and II, delimited by vertical red lines in Figure 3. The strongest emitters appear to be the edges of granules, at which the continuum intensity is a few percent less than the mean, i.e., brightness region I. The weakest emission comes from regions a few percent brighter than the mean, i.e., brightness region II.

In quantifying these variations in terms of realistic seismic emmissivity, it has to be borne in mind that a significant part of the egression-power signature represents noise emitted in the pupil of the helioseismic computations, not just from the focus. For this analysis, we suppose that the contribution of the noise from the pupil is equal to the contribution from the focus. This can be roughly expected if the acoustic power emitted downward from a local source, into the solar interior to re-emerge into a surface pupil some distance away, is matched by that emitted upward, however unwelcomely, directly into a pupil whose focus is some distant location. Under this supposition, the acoustic power emanating from brigtness region I in Figure 3 is about 2.5 times that emanating in brightness region II.

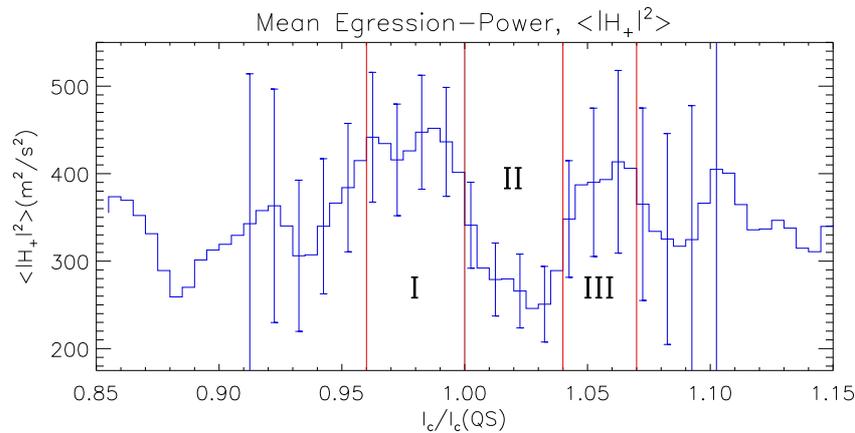

**Figure 3.** Egression power in the 5.5–6.5-mHz spectrum averaged over 0.005-unit intervals in normalized continuum brightness, as resolved by HMI, of regions from which seismic emission thereby represented emanates is plotted over the 0.85–1.15 range of normalized continuum brightness. Error bars indicate the standard statistical deviation of the egression power in respective 0.005-unit intervals of normalized continuum brightness.

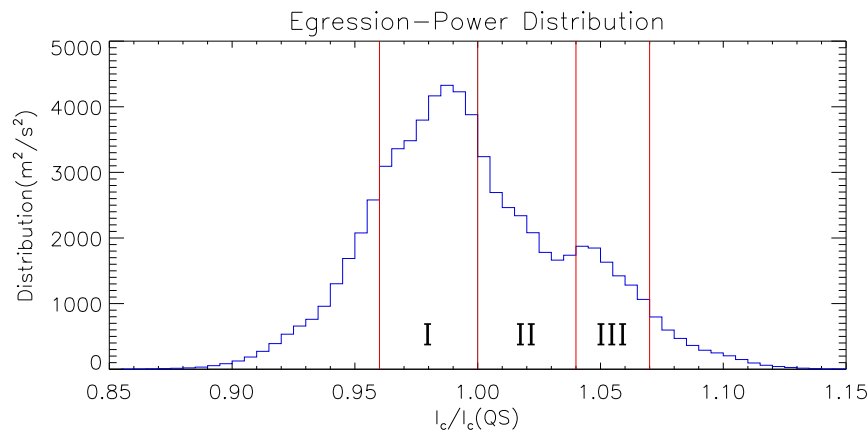

**Figure 4.** Statistical distribution of 5.5–6.5-mHz seismic source power is plotted over the brightness range 0.85–1.15 in normalized continuum intensity as resolved by HMI. This profile is scaled so that its integral over the unitless abscissa is the mean egression power.

The foregoing dependence may very well be expected based upon models that attribute seismic emission largely to down-flowing plumes [12] at the edges of granules. This might lead us to expect a maximum mean egression power at brightnesses that are intermediate between the bright granular interiors and the dark integranular lanes [14]. The mean egression power would then supposedly decrease from this relative maximum to reduced levels for brightnesses significantly greater, or less, than that of the maximum. This appears to be the case proceeding from left to right in Figure 3 through regions I and II, the maximum occuring in region I and dropping precipitously in region II. However, there appears to be a brightness region III, 4–7% brighter than the mean, of relatively strong seismic emission. The existence and placement of this feature brightward of the least emissive region, then, is not anticipated in the context of the rough understanding formulated above.

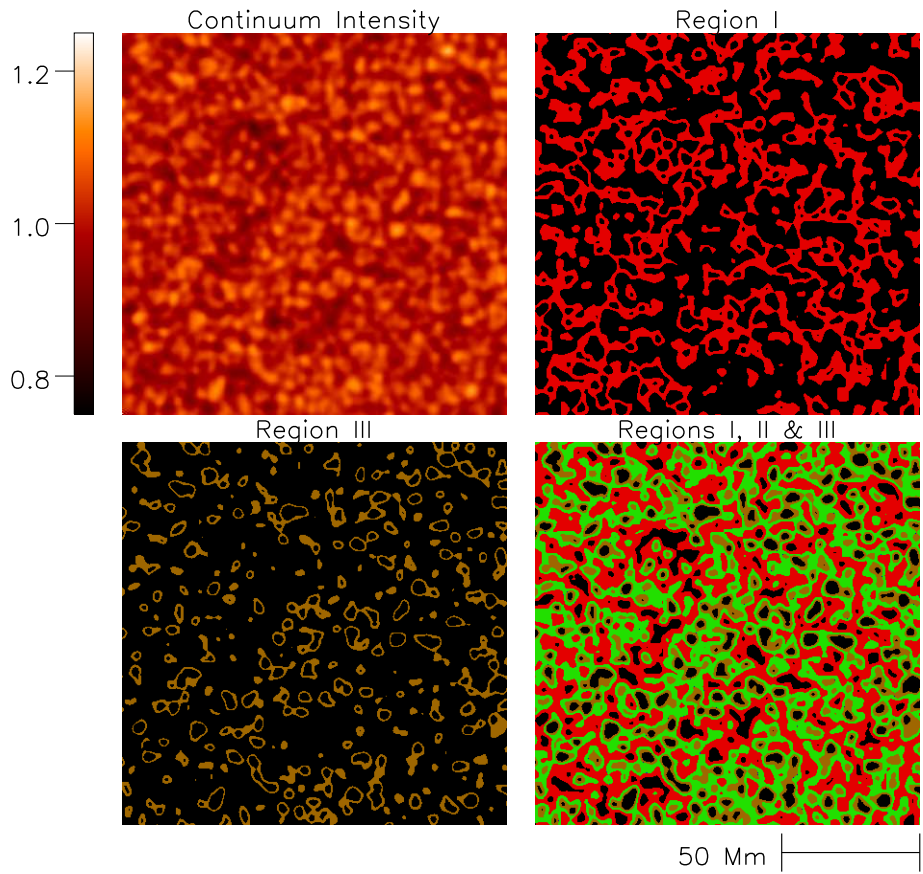

**Figure 5.** Regions whose continuum brightnesses (upper left) are in ranges in which 6-mHz egression power is statistically enhanced are mapped in the upper right (region I, 0–5% dimmer than the mean) and in the lower left (region III, 4–7% brighter than the mean) frames. Lower-right frame shows a composite map that renders in green the intervening range (0–4% brighter than the mean), in which acoustic emissivity is relatively suppressed.

Noting rapidly inflating error in Figure 3 for intensities more than 8% greater or less than unity, we forebear to comment on other interesting features of the profile plotted in Figure 3, except perhaps to note that, because of a rapidly decreasing distribution in the brightness histogram at brightnesses outside of $1.0 \pm 0.08$, the amount of seismic source power represented by these is not very large, while the statistical weight of the mean egression power diminishes, as indicated by the inflating error bars in Figure 3. The distribution in source power is shown by Figure 4. This profile is the product (see caption) of the profiles plotted in Figures 2 and 3.

## 3. Representation of Regions of Enhanced Seismic Source Power

Figure 4 identifies the brightness regions of anomalously high seismic emissivity in the upper right and lower left frames. The upper-left frame reproduces the continuum brightness rendered in the left frame of Figure 1. The upper-right frame renders region I as identified in Figures 3 and 4 in red, i.e., the brightness region of maximum mean egression power in Figure 3, whose continuum brightness is less than the mean. The lower left frame renders region III in light brown, i.e., the also-highly-acoustically-emissive region that is brighter than the mean. The lower-right frame shows a composite map that renders in green the intervening region, i.e., region II, whose brightness is greater than the mean but whose acoustic emissivity is relatively suppressed.

## 4. Discussion and Conclusions

The strong dependence of seismic source-power density upon continuum brightness as related to the structure of the solar granulation strongly supports the solar granulation as a major source of the solar oscillations. This dependence can be broadly summarized as follows:

(i) There is a significant correlation between the locations of the strongest episodes of high-frequency seismic source power and the edges of solar granules.
(ii) Statistically, the most seismically emissive regions are the edges of granules, where the continuum brightness is 0.95–0.99 times the quiet-Sun mean.
(iii) The least seismically emissive regions statistically are those whose continuum brightness is in the range 1.00–1.05 times the quiet-Sun mean.
(iv) However, regions in which the continuum brightness is 1.05–1.10 times the quiet-Sun mean, while less in area than the other two, are conspicuously emissive.
(v) The foregoing points are consistent with enhanced seismic emission in the neighborhoods of edges of photospheric granules. However, the significance of strong seismic emission from regions brighter than those from which the weakest seismic emission emanates, is somewhat of a mystery at this moment.

The last point should be moderated by the admission that the expectations challenged by this finding are based upon a relatively crude and highly qualitative understanding of how seismic emission from the solar granulation works. However, this result appears to invite some attention to aspects of p-mode generation that would lead to such a relationship. Some insight into this behavior might by acquired by applying the diagnostics developed in this study to simulations of the solar granulation such as those described by Skartlien, Stein & Nordlund [13].

### 4.1. Acknowledgements

We have have benefited deeply from the keen insights of Bob Stein and Mark Rast. This work has been supported by a contract from the Solar Heliospheric Physics program of the National Aeronautics and Space Administration.